\documentclass[aip,reprint,amsmath,amssymb,preprintnumbers]{revtex4-1}

\usepackage[utf8]{inputenc}
\usepackage[T1]{fontenc}
\usepackage{mathptmx}
\usepackage{etoolbox}
\usepackage{dcolumn}
\usepackage{bm}

\usepackage{color}
\usepackage{graphicx}
\usepackage{subcaption}
\usepackage{hyperref}
\usepackage[normalem]{ulem}
\usepackage{acro}

\date{\today}

\DeclareAcronym{OS}{short=OS, long=one-shift only}
\DeclareAcronym{SR}{short=SR, long=stochastic reconfiguration}
\DeclareAcronym{SR+LS}{short=SR+LS, long=stochastic reconfiguration with line search}
\DeclareAcronym{FN}{short=FN, long = fixed-node}
\DeclareAcronym{VMC}{short=VMC, long=variational Monte Carlo}
\DeclareAcronym{DMC}{short=DMC, long = diffusion Monte Carlo}
\DeclareAcronym{OO}{short=OO, long = orbital optimization}
\DeclareAcronym{SDJ}{short=SDJ, long = Slater determinant with Jastrow}
\DeclareAcronym{SPOs}{short=SPOs, long = single particle orbitals}
\DeclareAcronym{DFT}{short=DFT, long = density functional theory}
\DeclareAcronym{QMC}{short=QMC, long = quantum Monte Carlo}
\DeclareAcronym{CI}{short=CI, long = configuration interaction}

\begin{document}
\preprint{\large SAND2026-16112O }


\title[Assessing Orbital Optimization in Variational and Diffusion Monte Carlo]{Assessing Orbital Optimization in Variational and Diffusion Monte Carlo}

\author{Cody A. Melton}
\affiliation{Sandia National Laboratories, Albuquerque, NM 87125}
\email{cmelton@sandia.gov}

\author{Jaron T. Krogel}
\affiliation{Oak Ridge National Laboratory, Oak Ridge, TN 37830}
\email{krogeljt@ornl.gov}


\begin{abstract}
In this work, we investigate the fidelity of orbital optimization in variational Monte Carlo to improve diffusion Monte Carlo results on correlated magnetic systems, using CrSBr as a model system.
We compare the performance of different optimization methods, showing that stochastic reconfiguration is a robust and reliable optimizer.
We show that short range Jastrow factors are important for improving diffusion Monte Carlo, regardless of the quality of orbitals. 
Large active spaces are required to converge the variational energy, but ulitmately orbital optimization produces worse diffusion Monte Carlo energies when compared to standard orbitals from density functional theory. 
We show that this increased bias is due to larger locality errors from the use of pseudopotentials, while the fixed-node error is actually improved by using orbital optimization. 
Additionally, for observables other than energy, orbital optimization produces a systematically smaller mixed-estimator bias. 
Ultimately, we believe orbital optimization provides a reliable method to improve variational and pure fixed-node energies as well as lower mixed-estimator bias. 
\end{abstract}

\maketitle

Notice: This manuscript has been authored by UT-Battelle, LLC under Contract No. 
DE-AC05-00OR22725 with the U.S. Department of Energy. The United States 
Government retains and the publisher, by accepting the article for publication, 
acknowledges that the United States Government retains a non-exclusive, 
paid-up, irrevocable, worldwide license to publish or reproduce the published 
form of this manuscript, or allow others to do so, for United States Government 
purposes. The Department of Energy will provide public access to these results 
of federally sponsored research in accordance with the DOE Public Access Plan 
(http://energy.gov/downloads/doe-public-access-plan).

\section{Introduction}

A significant challenge in computational condensed matter physics is to accurately predict the properties of real materials. 
By far the most popular approach is \ac{DFT}\cite{jones_density_2015}, which has been extremely successful in describing weakly correlated systems. 
However, strong correlation and its influence on charge, spin, lattice, and topological degrees of freedom impedes an accurate \ac{DFT} description without empirical guidance from experiment, i.e. using experimental data to help select the ``best" functional.  
Explicit many-body methods such as \ac{QMC}\cite{foulkes_quantum_2001, luchow_quantum_2011} are well-equipped to capture these subtle effects and provide a more accurate description of the system from first principles, without relying on empirical guidance. 
While standard \ac{QMC} workflows have been largely successful \cite{shulenburger_quantum_2013, dubecky_quantum_2014, annaberdiyev_cohesion_2021, wines_toward_2025}, the fact that these methods are variational implies that they should be systematically improvable given a particular wave function ansatz. 
For example, moving toward increasingly larger multi-determinant wave functions which approach the full \ac{CI} limit show significantly improved energetics at both the \ac{VMC} and \ac{DMC} level, in both molecular and condensed systems\cite{caffarel_using_2016, malone_systematic_2020, shin_systematic_2024, spanedda_multireference_2025}.
Although multi-determinant wave functions have been demonstrated for condensed systems, they do not scale favorably with system size and therefore make it difficult for standard thermodynamic extrapolations as well as large defect systems. 
Therefore, it remains desirable to obtain the best single determinant wave function which scales favorably to hundreds of atoms. 

The largest source of error in \ac{DMC} comes from the fixed-node approximation\cite{benali_toward_2020}, which is controlled by the accuracy of the nodal surface of the underlying trial wave function. 
Historically, the fixed-node error has been minimized by calculating energetics using a variety of wave functions built from \ac{DFT} orbitals and selecting the set that yields the lowest \ac{DMC} energy. 
This can be achieved by calculating a broad range of functionals, or by scanning single parameter functionals such as \ac{DFT}+U or hybrid functional exact exchange mixing fractions\cite{kolorenc_wave_2010,luo_phase_2016}. 
While this does provide the lowest \ac{DMC} energy over the orbital sets considered, this approach does not provide a systematic procedure to guide us to the optimal single particle orbital set. 
Coupling direct \ac{DMC} into an optimization scheme where the free parameters of the wave function are directly optimized would allow for systematic improvement toward the optimal nodal surface, but in general this is too expensive. 

Direct optimization of the wave function at the \ac{VMC} level is most often used to optimize Jastrow parameters. 
However, in principle any parameter in the wave function can be optimized, including those that directly influence the nodal surface. 
For example, full optimization of the Jastrow, determinant expansion parameters, orbital rotation parameters, and underlying orbital basis set parameters can all be optimized variationally\cite{toulouse_optimization_2007, toulouse_full_2008}. 
If we limit ourselves to single determinant wave functions, which is of interest for condensed system where more general ansatze are still prohibitive, we have the freedom to optimize rotations between our occupied and virtual orbitals as well as potentially optimizing the underlying basis functions. 
Directly optimizing the orbitals through orbital rotations, which we will denote throughout this paper denote as \ac{OO}, is the simplest approach to modifying the nodal surface at the \ac{VMC} level and potentially yield improved nodal surfaces within \ac{DMC}. 
Previously, this has been shown to provide starting point independence from the underlying \ac{DFT} orbital set \cite{townsend_starting-point-independent_2020}
, however there is evidence that \ac{OO} alone may not improve the nodal surface even if the \ac{VMC} energy is improved\cite{chow_capturing_2024}.


A class of materials where systematically improvable wave functions in \ac{QMC} could have a significant impact is correlated magnetic systems. 
For example, current \ac{DFT} functional approximations indicate an ``inverse Jacob's ladder" when predicting the magnetic moments of bulk ferromagnetic transition metal materials \cite{fu_density_2019}, where more advanced functionals are less accurate than the simplest local density approximation.
For 2D materials, where electron correlation can be enhanced by the reduced dimensionality, mean-field descriptions of the electronic structure can also be insufficient. 
For example, in CrI$_3$ different functionals predict a wide range of lattice structures, magnetic moments, and even the physical origin of magnetism \cite{staros_combined_2022}.
Given the potential technological impact of magnetic 2D materials, ranging from spintronics to quantum information \cite{zhang_2d_2024}, many-body methods such as \ac{QMC} will become increasingly valuable. 

In this work, we apply \ac{OO} as an approach to improve the wave function on CrSBr, a semiconducting 2D van der Waals material. 
CrSBr has recently garnered a lot of attention due to high critical temperature and air stability at ambient conditions, which makes it an attractive candidate for experiment \cite{ziebel_crsbr_2024}. 
 We investigate the fidelity of \ac{OO} in \ac{VMC} and its impact on \ac{DMC} from a variety of fronts for CrSBr. 
We consider the influence of different optimization procedures, each of which have slightly different cost functions, and compare its impact to both \ac{VMC} and \ac{DMC}. 
We also consider the coupling between orbital rotation parameters and varying fidelity Jastrow factors. 
Another consideration is how many virtual orbitals (active space) are necessary in order to converge the energy at both the \ac{VMC} and \ac{DMC} level using the Kohn-Sham orbitals we consider here.

Ultimately, we find that \ac{OO} carried out at the \ac{VMC} level is not able to produce the lowest \ac{DMC} energy when compared against scans using \ac{DFT}+U orbitals, even when considering large active spaces for the orbital rotations. 
This is not unanticipated since theoretically there is no guarantee that improved variational wave functions within a restricted and approximate ansatz will yield improved \ac{DMC} energies. 
Whether or not this is true in practice for correlated materials is an important question to answer. 
We show that \ac{OO} is marginally improving the nodal surface, and therefore reducing the fixed node error. 
However, the \ac{DMC} energies are higher due to an increased magnitude in the locality error, which arise from the use of nonlocal pseudopotentials in the Hamiltonian. 
We also note that even though the \ac{DMC} energetics are marginally worse, for observables other than the energy we show that it is still be beneficial to use \ac{OO} as the significant improvement at the variational level tends to reduce mixed estimator biases. 
This increased accuracy is important because quantities like the electron density, for example, can be used to benchmark, and potentially improve, other theories such as DFT. 

This paper is organized as follows. 
In \S \ref{sec:methods}, we give a brief overview of the methods used in this work. 
We describe the Slater-Jastrow wave function ansatz we use, including a description of all parameters to be optimized. 
Additionally, we compare and contrast all of the optimization methods used in this work. 
In \S \ref{sec:system}, we describe the CrBS system we use to carry out the comparisons. 
In \S \ref{sec:oo_optimizers}, we compare the performance of different optimization methods available on \ac{OO}. 
We discuss the influence of Jastrow quality and how it influences the \ac{OO} procedure in \S \ref{sec:oo_jastrows}. 
We contrast the different convergence behavior of \ac{OO} with respect to active space size in both \ac{VMC} and \ac{DMC} in \S \ref{sec:oo_active_space}.
Lastly, we present our conclusions and future directions in \S \ref{sec:conclusions}.

\section{Methods} 
\label{sec:methods}

In this section we describe the methodology used in this work. 
All \ac{DFT} and \ac{DFT}+U orbital calculations are performed using \textsc{Quantum ESPRESSO} \cite{giannozzi_quantum_2009,giannozzi_advanced_2017}. 
All \ac{QMC} calculations are performed using \textsc{QMCPACK} \cite{kim_qmcpack_2018}.
We utilize existing optimization methods in \textsc{QMCPACK} such as the linear method using only one shift \ac{OS}, as well as the stochastic reconfiguration \ac{SR} method\cite{sorella_generalized_2001, casula_correlated_2004}.
As we will show, \ac{SR} approaches are better suited for large parameter count optimizations. 

\subsection{Wavefunction Ansatz}

The Slater-Jastrow wave function is the most commonly used ansatz in the \ac{QMC} community. 
Here we extend the Slater-Jastrow wave function to include orbital optimization through the use of a unitary operator $e^{\hat{\kappa}_\sigma}$, namely
\begin{equation}
    \Psi_T(\mathbf{R}) = e^{-J(\mathbf{R})} \prod_{\sigma} e^{\hat{\kappa}_\sigma}\det\left[\phi^\sigma_i(\mathbf{r}_{j,\sigma})\right],
\end{equation}
where $J(\mathbf{R})$ is the Jastrow factor and the antisymmetric part is formed out of Slater determinants of \ac{SPOs} coming from \ac{DFT} calculations, represented as a matrix $\phi_i^\sigma(\mathbf{r}_{j,\sigma})$ with orbital index $i$, electron index $j$ and spin index $\sigma$. 
The Jastrow factor can be formally expanded over $n$-body interactions
\begin{equation}
    J = \sum_{i,\sigma} u_1(\mathbf{r}_{i,\sigma}) + \sum_{i,j,\sigma,\mu} u_2(\mathbf{r}_{i,\sigma}, \mathbf{r}_{j,\mu}) + \sum_{i,j,k}u_3(\mathbf{r}_i, \mathbf{r}_j, \mathbf{r}_k) + \ldots
\end{equation}
 
In \textsc{QMCPACK}, the  expansion is truncated at $u_2$, including short-ranged electron-ion and electron-electron-ion inhomogeneous terms.  The Kato cusp conditions\cite{kato_eigenfunctions_1957} are incorporated into the homogeneous part of the two-body function.
We may optionally include some long-ranged Jastrow terms which are important in the smallest simulation cells, which we call $k$-space Jastrows.  
A general two-body $k$-space Jastrow\cite{drummond_jastrow_2004} is typically written in terms of the structure factor
\begin{equation}
    J^{ab}_{k} = \sum_{\mathbf{G}\ne 0}a_\mathbf{G} \rho^a_\mathbf{G} \rho^b_{-\mathbf{G}},
\end{equation}
where $\mathbf{G}$ are the standard $\mathbf{G}$-vectors of the reciprocal lattice and $a_\mathbf{G}$ are the optimizable coefficients. 
The density operator for species $a$ in reciprocal space is given by 
\begin{equation}
    \rho^a_\mathbf{G} = \sum_i^{N_a} e^{i \mathbf{G}\cdot \mathbf{r}_{i}}.
\end{equation}
In this work we will consider 1-body $k$-space Jastrows between electrons and ions, and 2-body $k$-space Jastrows between electrons only.
The $\mathbf{G}$-vector cutoff in reciprocal space is always chosen as $G_c = \left(6 N_e  \pi^2 /V\right)^{1/3}$, where $N_e$ is the number of electrons and $V$ is the volume of the simulation cell. 

The orbital optimization is accomplished by optimizing the parameters in the unitary operator $e^{\hat{\kappa}_\sigma}$, where
\begin{equation}
    \hat{\kappa}_\sigma = \sum\limits_{i,j} \kappa^\sigma_{ji} \left( \hat{a}^\dagger_{i,\sigma} \hat{a}_{j,\sigma}  - \hat{a}^\dagger_{j,\sigma} \hat{a}_{i,\sigma}\right)
\end{equation}
where the $\hat{a}^\dagger_{i,\sigma},\hat{a}_{i,\sigma}$ operators are the creation/annihilation operators for spin orbital $(i,\sigma)$ \cite{toulouse_optimization_2007,toulouse_full_2008}.  
The $\kappa^\sigma_{ji}$ is an anti-hermitian matrix for spin-channel $\sigma$, whose elements are the tunable parameters. 
For this work, we only consider rotations the orbital set we obtain from our underlying DFT set of orbitals, i.e. the Kohn-Sham occupied and virtual orbitals. 
The variational freedom can be increased by enlarging the active space by adding more virtual orbitals.


\subsection{Optimizers}

Optimization of the wave function in \ac{QMC} methods have historically relied on two major approaches, namely the linear method \cite{umrigar_alleviation_2007}  and \ac{SR} \cite{sorella_generalized_2001, casula_correlated_2004}.
We briefly summarize the differences in the optimization approaches available within QMCPACK. 
We also note that throughout this work, all optimizations steps use approximately 5$\times 10^6$ samples per optimization iteration in order to sufficiently resolve the variational energy at each step of the optimization.

\subsubsection{One-Shift Only Optimizer}
The One-Shift (\ac{OS}) optimizer in QMCPACK is based on the linear method \cite{toulouse_optimization_2007, umrigar_alleviation_2007}, which is historically the most popular approach used in real space \ac{QMC}. 
The linear method works by expanding the proposed wave function $|\Psi'\rangle$ in the space of parameter derivatives around the current wave function $|\Psi_0\rangle$, 
\begin{equation}
    |\Psi'\rangle = |\Psi_0\rangle + \sum_{i=1}^{N_p}\Delta \theta_i \left| \Psi_{\theta_i}\right\rangle,
    \label{eqn:wf_expansion}
\end{equation}
where $\Delta \boldsymbol{\theta} = \left\{ \Delta \theta_i \right\}_{i=1}^{N_p}$ is the vector of $N_p$ variational parameters and $|\Psi_{\theta_i}\rangle = \partial_{\theta_i} |\Psi_0\rangle$ are the vector of parameter derivatives of the wave function. 
Minimizing the variational energy of the new wave function leads to an equation for the parameter update given by a generalized eigenvalue problem
\begin{equation}
    \boldsymbol{H} \cdot \Delta \boldsymbol{\theta} = E \boldsymbol{S} \cdot \Delta \boldsymbol{\theta}
\end{equation}
where $\Delta\theta$ comprises a potential parameter update.
The \ac{OS} optimizer accepts or rejects the parameter update based on the magnitude of the wavefunction ratio found via correlated sampling, with small ratios resulting in rejection.
As this is the most common optimizer used for typical Jastrow optimizations in \textsc{QMCPACK}, we will focus only on this variant of the linear method. 

\subsubsection{Stochastic Reconfiguration} 
While the \ac{OS} is based on the linear method, which is a quasi-Newton scheme, another popular optimization method related to gradient descent methods is \ac{SR} \cite{sorella_green_1998, sorella_generalized_2001}.  
A simple formulation of SR updates the wavefunction based on a short time projection:
\begin{eqnarray}
 |\Psi'\rangle = e^{-\tau \hat{H}}|\Psi_0\rangle \approx (1-\tau\hat{H})|\Psi_0\rangle
\end{eqnarray}
Expanding in wave function derivatives (Eq. \eqref{eqn:wf_expansion}) gives a linear equation for the parameter update
\begin{eqnarray}
    \mathbf{S}\Delta \boldsymbol{\theta} = -\tau \mathbf{h}
    \label{eqn:sr_update}
\end{eqnarray}
This optimizer has the benefit that parameter derivatives of the Hamiltonian do not need to be evaluated, only parameter derivatives of the wave function. 
\ac{SR} methods have become popular in the machine learning wave function community since \ac{SR} has favorable scaling with parameters relative to linear methods \cite{chen_empowering_2024,goldshlager_kaczmarz-inspired_2024,armegioiu_functional_2025}. 
In pure \ac{SR}, we directly accept the parameter updates from Eq. \eqref{eqn:sr_update}. 
As a potential way to accelerate convergence, a correlated sampling line search can be performed \cite{wheeler_pyqmc_2023}.
The cost function minimized along the line search can be chosen as a linear combination of energy and variance. 
We refer to this optimizer throughout as \ac{SR+LS}.

\subsection{DMC} 
The details of \ac{DMC} have been covered extensively elsewhere \cite{foulkes_quantum_2001, luchow_quantum_2011, kim_qmcpack_2018}, so we briefly describe it here. 
The accuracy of \ac{DMC} comes from the imaginary time propagator, where the ground state wave function is projected out from some initial trial wave function in the long imaginary time limit, i.e. $|\Phi_0\rangle = \lim\limits_{\tau\rightarrow\infty}\exp\left[ -\tau(\hat{H} - E_0)\right] |\Phi\rangle$. 
To avoid the fermion sign problem, we use the fixed node approximation in which the nodes of the projected wavefunction are constrained to match those of the trial wavefunction.
This approximation is often assumed to be the largest source of bias in DMC calculations.

Since almost all practical calculations utilize pseudopotentials to remove inert core electrons, there is an additional sign problem in the projector since the nonlocal pseudopotential matrix elements  can be positive or negative. 
The sign problem can be avoided if we simply ignore the part of the Green's function arising from the nonlocal pseudopotential, known as the locality approximation \cite{mitas_nonlocal_1991}, but as a result the variational principle is lost. 
In the alternative ``T-moves'' \cite{casula_beyond_2006, casula_size-consistent_2010} approach,  the variational principle is restored but with added bias due to neglect of the sign-changing terms.
The quality of the Jastrow factors directly impact the magnitude of the locality error, and previous work has shown that the pure fixed node energy can be determined via extrapolations with respect to the \ac{VMC} and \ac{DMC} energies as the Jastrow factor is varied, thus also providing the magnitude of the locality error\cite{krogel_magnitude_2017}.

A typical strategy to find the ``best" Slater-Jastrow wavefunction
is to run DMC with a few different functionals, or perform a 1D scan over parameters such as the Hubbard $U$ or exact exchange mixing fraction. 
This optimization uses the variational \ac{DMC} energy as the cost function, but only explores a fraction of the variational freedom allowed within the Slater-Jastrow ansatz
However, the variational principle also applies to \ac{VMC}, permitting wave function optimization at significantly lower cost. 
If the Jastrow factor is optimized, for example, the nodal surface remains unchanged, but the magnitude of the locality error is impacted. 
By optimizing parameters that directly modify the nodal surface alongside the Jastrows in \ac{OO}, we impact both the DMC fixed-node and locality errors.
In this work, we are interested in determining the extent to which VMC orbital optimization results in reductions of errors at the \ac{DMC} level, which we explore in detail below.

\section{Results}
\label{sec:results}

\subsection{CrSBr}
\label{sec:system}
\begin{figure}
    \centering
    \includegraphics[width=\linewidth]{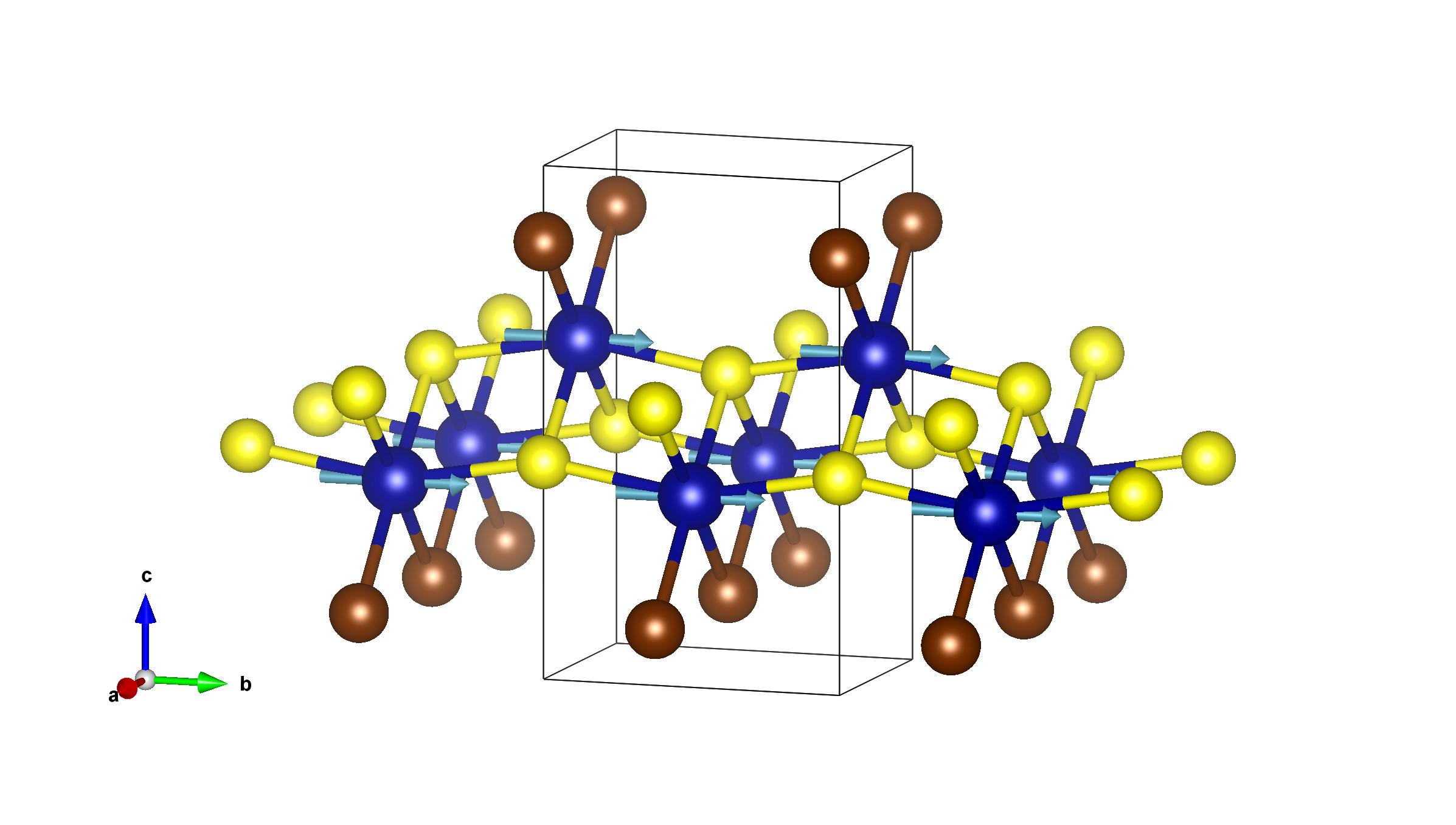}
    \caption{Structure of the quasi-2D CrSBr system in a ferromagnetic configuration, made with \textsc{VESTA} \cite{momma_vesta_2011}. The magnetic moments on the chromium atoms are also indicated. }
    \label{fig:structure}
\end{figure}

In this work we investigate the performance of \ac{OO} in \ac{VMC} and \ac{DMC} in the context of the correlated 2D semiconductor CrSBr. 
The ground state magnetic configuration is antiferromagnetic between each layer and ferromagnetic within a layer, with the magnetic moment lying along the $b$-axis\cite{lee_magnetic_2021,telford_layered_2020}.
The structure, including the magnetic moments, is shown in Fig. \ref{fig:structure}. 
Neutron scattering experiments indicate a saturation moment of $\sim 3 \mu_B$ per Cr site\cite{scheie_spin_2022}. 
Since we are currently interested in determining whether or not \ac{OO}  within \ac{VMC} also improves properties in \ac{DMC}, we limit ourselves to the primitive cell with a single layer and do not consider finite-size effects for simplicity. 
This restricts CrSBr to the FM phase, but since we consider independent orbital rotations for the up and down orbitals, 
the performance of OO we examine here should also extend to arbitrary collinear magnetic order.
For the \ac{DFT}+U calculations, we consider Hubbard parameters between 0-5~eV starting with an initial magnetic moment of $3\mu_B$ per Cr. 
In order to probe \ac{OO} for the general case, we consider a single random $k$-point such that the wave function is complex. 

We first consider the impact of different optimization methods in \textsc{QMCPACK}, to determine if one is to be preferred over another. 
Next, we assess how increasing the sophistication of Jastrow factors impacts the quality of the orbitally optimized wavefunction at both the VMC and DMC levels.
We then explore the convergence of \ac{OO} with respect to the total number of virtual orbitals considered in the optimization. 
Following this, we disentangle the impact of OO on DMC locality and fixed node errors to allow a clearer comparison between OO and the traditional ``U-scan''. To the best of our knowledge, this is the first time a detailed analysis constrasting these two errors has been made in the context of OO.
Lastly, we show how performing \ac{OO} influences the quality of DMC observables other than the energy. 

\subsection{Orbital optimization with different optimizers}
\label{sec:oo_optimizers}

\begin{figure*}
  \centering
  \begin{subfigure}[b]{0.48\textwidth}
    \centering
    \includegraphics[width=\linewidth]{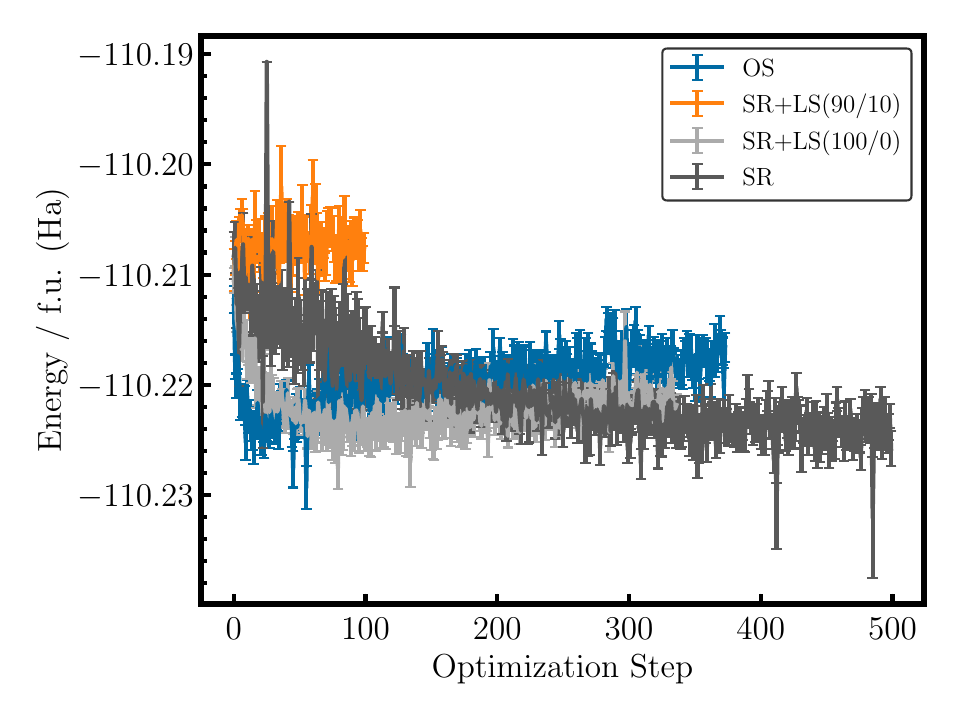}
    \caption{Energy of \ac{OO} with 37 total orbitals}
    \label{fig:oo_n37}
  \end{subfigure}
  \hfill
  \begin{subfigure}[b]{0.48\textwidth}
    \centering
    \includegraphics[width=\linewidth]{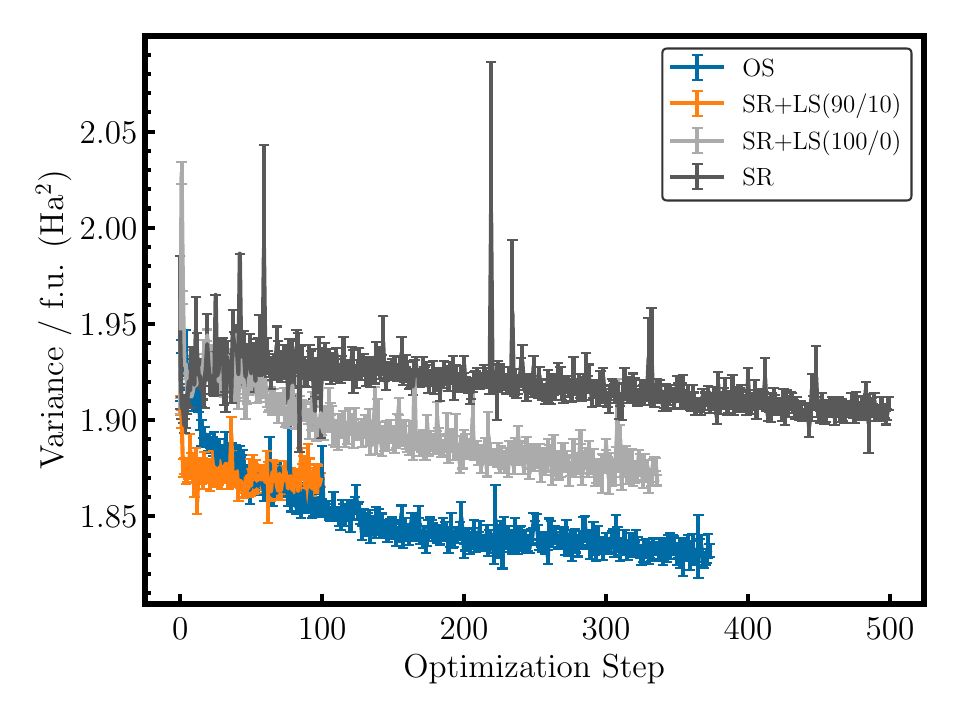}
    \caption{Variance of \ac{OO} with 37 orbitals}
    \label{fig:oo_n37_var}
  \end{subfigure}
  \begin{subfigure}[b]{0.48\textwidth}
    \centering
    \includegraphics[width=\linewidth]{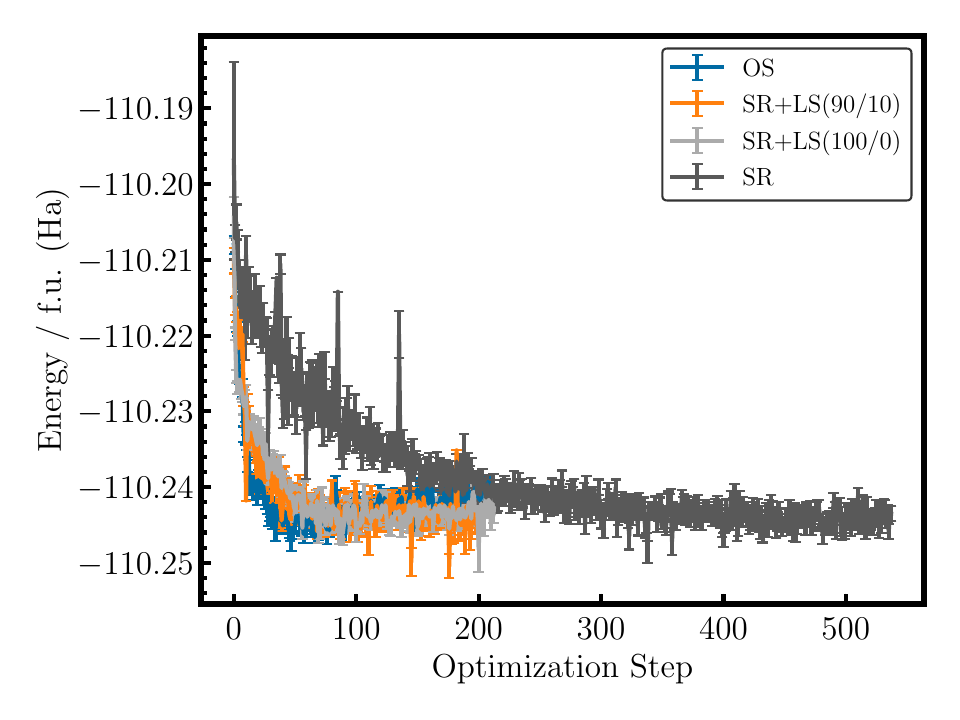}
    \caption{Energy of \ac{OO} with 60 total orbitals}
    \label{fig:oo_n60}
  \end{subfigure}
  \begin{subfigure}[b]{0.48\textwidth}
    \centering
    \includegraphics[width=\linewidth]{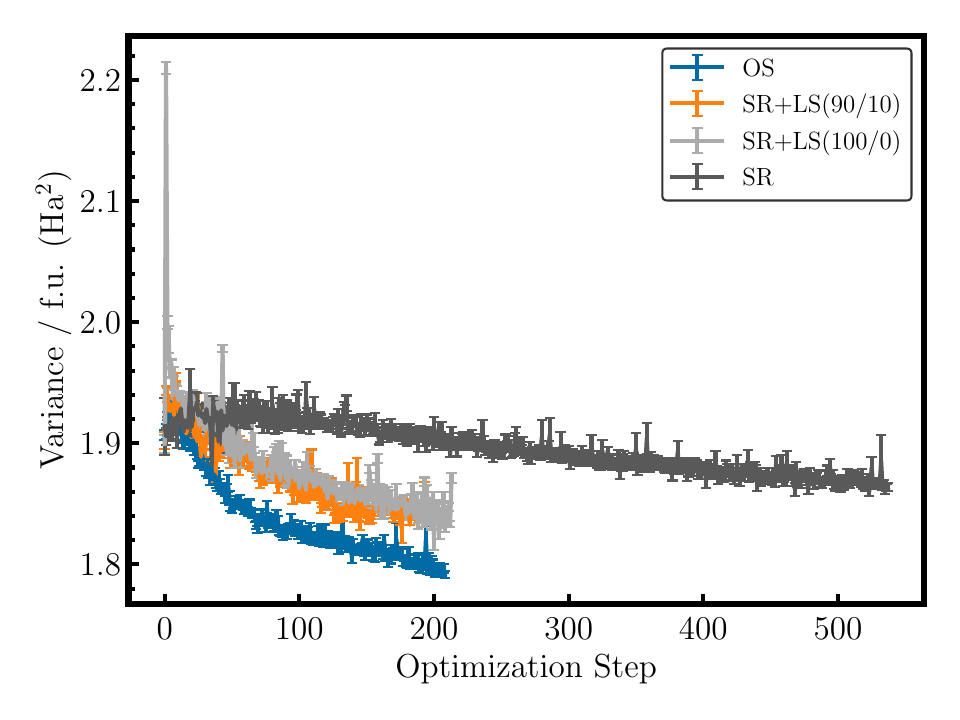}
    \caption{Variance of \ac{OO} with 60 total orbitals}
    \label{fig:oo_n60_var}
  \end{subfigure}
  \caption{Orbital optimization using different optimization strategies with different active spaces, (\subref{fig:oo_n37} \& \subref{fig:oo_n37_var}) with 37 orbitals and (\subref{fig:oo_n60} \& \subref{fig:oo_n60_var}) with 60 orbitals. }
  \label{fig:opt_compare}
\end{figure*}

In Figure \ref{fig:opt_compare}, we compare the various optimizers currently available in \textsc{QMCPACK}, namely the \ac{OS}, \ac{SR}, and \ac{SR+LS} optimizers. 
For the \ac{SR+LS} optimizer, the cost function in the correlated sampling line search step can be a linear combination of energy and variance, which we label as $(X/Y)$ to indicate $X \%$ energy and $Y \%$ variance.
For the smallest active space (Figure \ref{fig:opt_compare}a), we see that \ac{SR+LS}(90/10) (orange) does not actually lower the energy, whereas all other optimizers do. 
The \ac{OS} and \ac{SR+LS}(100/0) optimizers show a rapid initial decrease in energy, however the energy in each of these optimizers begins to increase over time. 
It should be noted that in both cases of \ac{OS} and \ac{SR+LS}, the optimizers rely on correlated sampling for parameter updates. 
In pure \ac{SR}, which does not rely on correlated sampling, the energy is consistently decreasing but with a slow convergence.

In a larger active space of 60 orbitals (Figure \ref{fig:oo_n60}), we see a similar trend where the \ac{OS} optimizer begins to drift toward higher energies. 
The various flavors of \ac{SR+LS} lower the energy rather quickly, and seem to converge by 200 iterations. 
Pure \ac{SR} also converges slowly, plateauing around 500 iterations. 
In each active space, the variance never plateaus and continuously decreases throughout the optimization. 
The fact the variances are so different across each optimizer suggests that each optimizer is minimizing a different objective function.
It is also interesting to note that the optimizers that rely on correlated sampling for the parameter update do not converge to the VMC energy minimum.

\begin{figure*}
  \centering
  \begin{subfigure}[b]{0.48\textwidth}
    \centering
    \includegraphics[width=\linewidth]{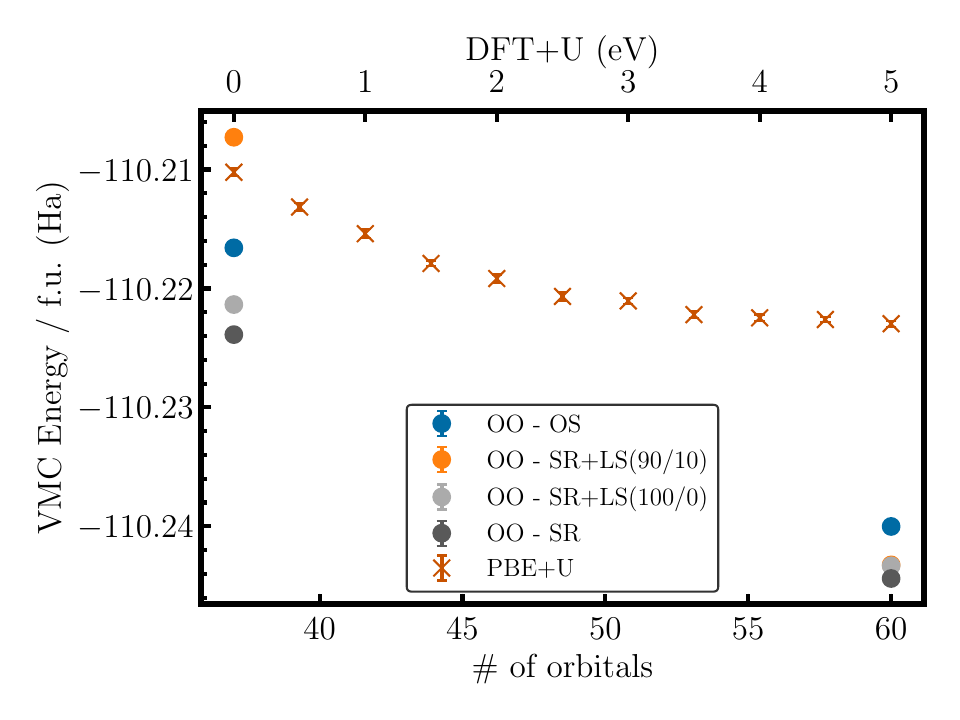}
    \caption{VMC}
    \label{fig:vmc_compare}
  \end{subfigure}
  \hfill
  \begin{subfigure}[b]{0.48\textwidth}
    \centering
    \includegraphics[width=\linewidth]{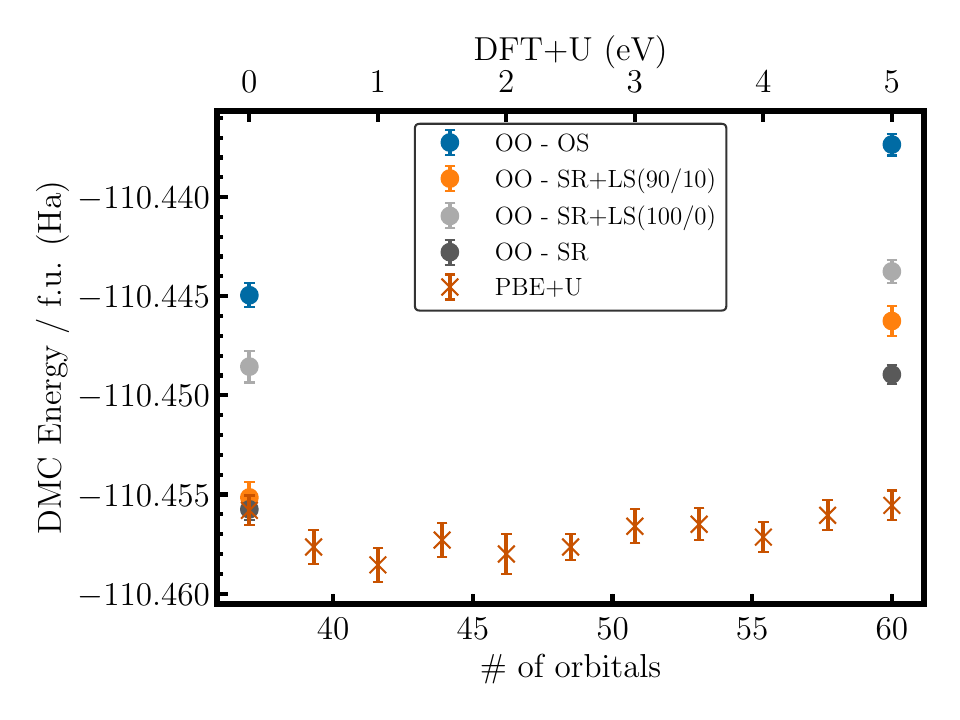}
    \caption{DMC}
    \label{fig:dmc_compare}
  \end{subfigure}
  \caption{Comparison of different optimization strategies in \ac{VMC} and \ac{DMC}. In PBE+U orbitals, only the jastrow is optimized and is plotted using the top $x$-axis, whereas the rest optimize the orbitals with different active space sizes and are shown across the bottom $x$-axis. Note that in (\subref{fig:vmc_compare}), all error bars are smaller than the points.}
  \label{fig:qmc_compare}
\end{figure*}

Ultimately, we are interested in improving the \ac{DMC} energies.
For transition metal bearing materials, a common method to ``optimize" the \ac{FN} bias generate to perform DFT+U/hybrid DFT for a variety of U/EXX parameters. 
For each of these orbital sets, the best wavefunction is selected by choosing the U/EXX which gives the lowest \ac{DMC} energy. 
While this approach directly minimizes the desired quantity, namely the \ac{DMC} energy, it can only vary the orbitals over a single parameter unlike the \ac{OO} approach we consider here. 

In Figure (\ref{fig:qmc_compare}), we compare the \ac{VMC} and \ac{DMC} energies for the different optimizers, including a scan of DFT+U orbitals as a baseline. 
In \ac{VMC}, we see that there is a clear decrease in the energy as a function of $U$.
However, by incorporating \ac{OO} the VMC energy can be significantly improved, and is robustly lower in energy than the \ac{DFT}+U scan for all values of U, even at the smallest active space using the \ac{SR} optimizer. 
Unfortunately, this improvement does not carry over into an improvement in \ac{DMC}, as seen in Figure (\ref{fig:dmc_compare}). 
Clearly, each optimizer has different convergence behavior and are finding different nodal surfaces, even if the variational energies are not significantly different, which is expected given the different variance convergence behavior. 
From these results at small active space size, the best nodal surface seemingly comes from doing a simple DFT+U scan instead of performing \ac{OO}. 
This system is also unusual in that we typically see a more clear minimum when doing the \ac{DFT}+U scan at the \ac{DMC} level \cite{luo_phase_2016,townsend_starting-point-independent_2020,staros_combined_2022,ahn_diffusion_2025}.  
Even though \ac{SR} shows a slightly worse \ac{DMC} energy, the \ac{VMC} energy is significantly improved over the DFT+U orbitals surface. 
Given this, and the fact that \ac{SR} is the only optimizer that produces a robust optimization where the energy consistently decreases, we use \ac{SR} for further optimization investigations below. 

\subsection{Impact of Jastrow factors on orbital optimization}
\label{sec:oo_jastrows}
In the previous section, we used a fixed Jastrow factor, namely a short-ranged 1-, 2- and 3- body (electron-electron-ion) Jastrow which is commonly used in standard QMC workflows. 
However, the VMC optimized orbitals should change as different functional forms of the Jastrow factor are included in the optimization.
In order to explore this, we carried out a variety of \ac{OO} calculations with different Jastrow forms. 
We considered the standard short-ranged 1- and 2- body Jastrow (J12), J12 with one-body k-space  (J12k1), two-body kspace (J12k2), and a 1- and 2- body kspace (J12k12) terms. 
We also considered those combinations for a short-ranged 1-, 2-, and 3- body Jastrow factor (J123). 
We initialized all rotation and Jastrow parameters to zero and independently optimized them for a 45 orbital active space.
Here, we compare wavefunctions comprised of fixed PBE orbitals and optimized Jastrow factors with ones where both the orbitals and the various Jastrow factors are optimized.

\begin{figure*}
    \centering
    \begin{subfigure}[b]{0.48\textwidth}
        \centering
        \includegraphics[width=\linewidth]{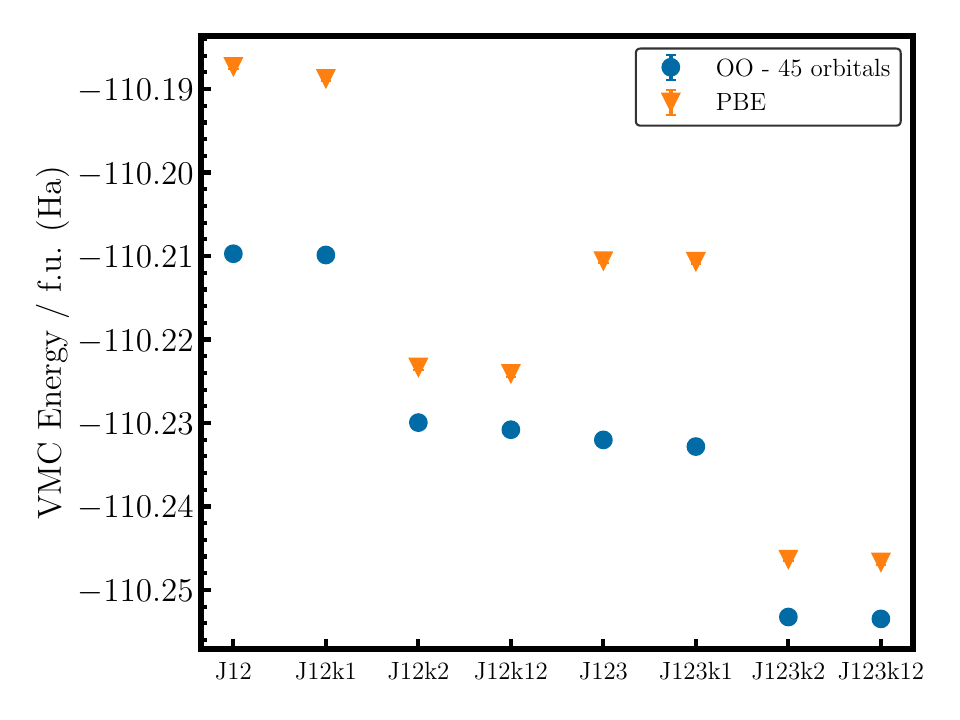}
        \caption{\ac{VMC}}
        \label{fig:jastrow_vmc}
    \end{subfigure}
    \hfill
    \begin{subfigure}[b]{0.48\textwidth}
        \centering
        \includegraphics[width=\linewidth]{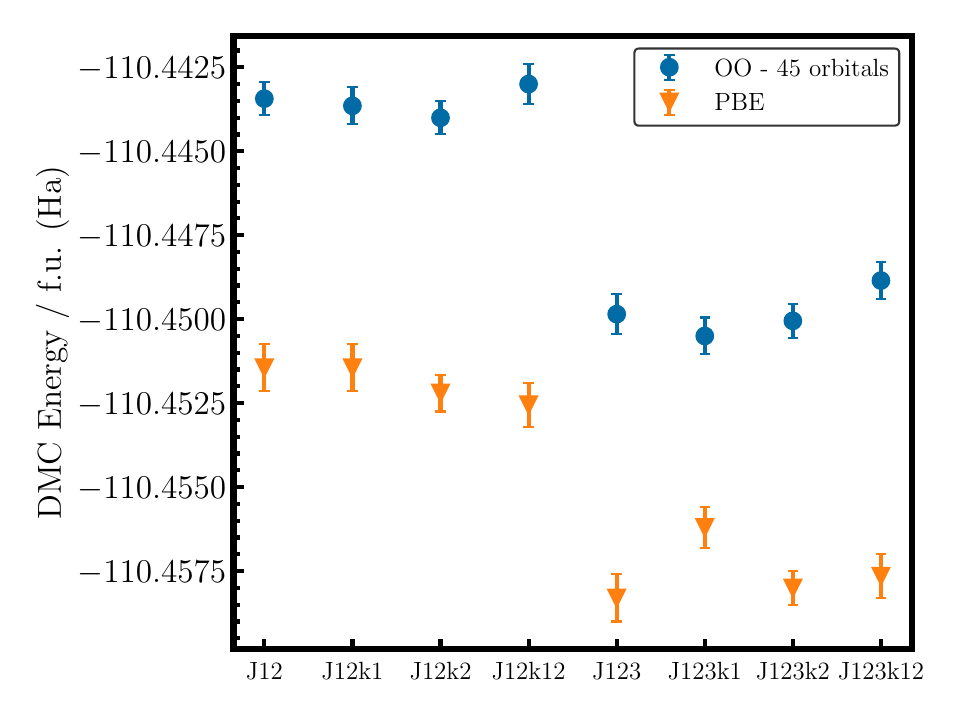}
        \caption{\ac{DMC}}
        \label{fig:jastrow_dmc}
    \end{subfigure}
    \caption{Comparison of \ac{OO} using different Jastrow factors, with \ac{VMC} shown in (\subref{fig:jastrow_vmc}) and \ac{DMC} shown in (\subref{fig:jastrow_dmc}). Note that in (\subref{fig:jastrow_vmc}), there errorbars are smaller than the points on this scale. }
    \label{fig:jastrow_scan}
\end{figure*}
The VMC and DMC energies of each of these wavefunctions is shown in Figure \ref{fig:jastrow_scan}. 
In all cases, we observe that improving the Jastrow ansatz recovers increasing amounts of variational energy at the VMC level (see Fig. \ref{fig:jastrow_vmc}).
The one-body k-space terms offer no improvement at the variational level for any set of orbitals, whereas two-body long range correlation significantly lowers the energy. 
With fixed orbitals, the two-body long range correlation significantly improves the energy, and even outperforms a three-body short range Jastrow. 
By allowing the k-space terms to be optimized alongside orbital rotation parameters there is a systematic improvement over corresponding fixed orbital sets. 
Therefore, \ac{OO} is able to recover more variational energy with all Jastrow types, although the gain in energy is smaller if long range correlations are included. 


A somewhat different picture emerges in the DMC results for the same wave functions (see Fig. \ref{fig:jastrow_dmc}).
For the orbitally optimized wave functions (blue), the long-ranged k-space terms do not affect the \ac{DMC} energies of the optimized wavefunctions at fixed short-ranged order (i.e. J12 and J123 separately).
However, a significant effect is seen as the sophistication of the short-ranged form is increased: better short-ranged correlation leads to improved \ac{DMC} energies. 
This suggests that further improvements to short-ranged Jastrow forms (see e.g. reference \cite{lopez_rios_framework_2012}) may provide a path toward optimal Slater-Jastrow \ac{DMC} energetics when coupled with \ac{OO}. 
The topic bears further investigation in future work by ourselves or others.

Within the limitations of the Jastrow forms we employ, we are unable to produce optimized orbitals with nodal structures matching or exceeding those found with DFT alone. 
In the next section, we additionally explore the role of increased active space size on the nodal structure using the J123 Jastrow form, since the inclusion long-range correlation does not impact the \ac{DMC} quality.

\subsection{Orbital Optimization vs. active space}
\label{sec:oo_active_space}

\begin{figure*}
  \centering
  \begin{subfigure}[b]{0.48\textwidth}
    \centering
    \includegraphics[width=\linewidth]{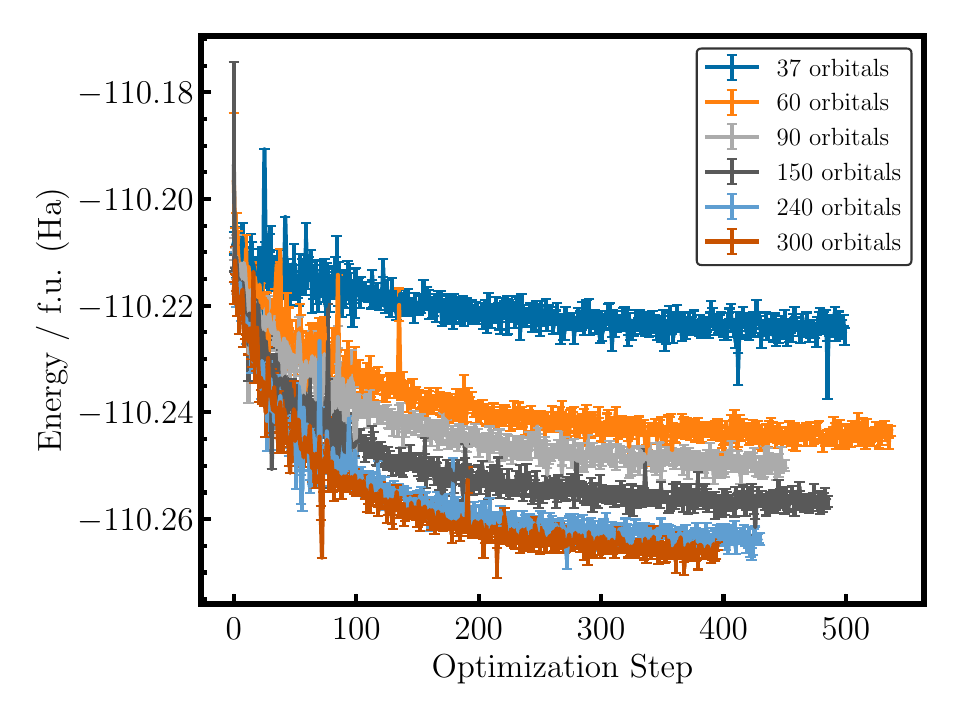}
    \caption{\ac{OO} energies vs. active space}
    \label{fig:act_energy}
  \end{subfigure}
  \hfill
  \begin{subfigure}[b]{0.48\textwidth}
    \centering
    \includegraphics[width=\linewidth]{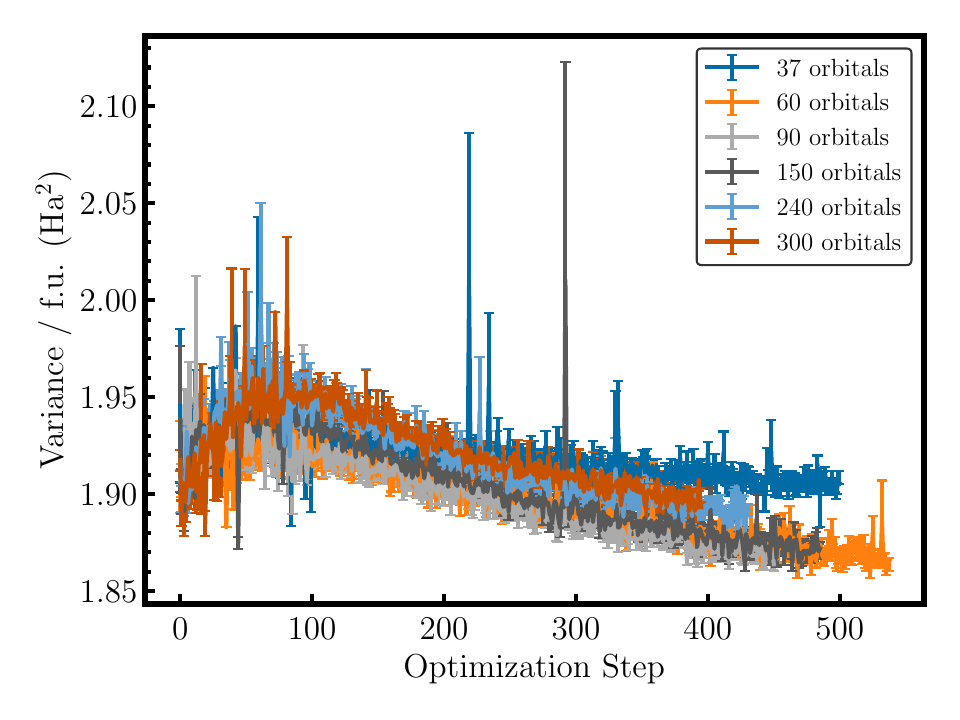}
    \caption{\ac{OO} variances vs. active space}
    \label{fig:act_variance}
  \end{subfigure}
  \caption{Comparison of optimizations with different active space sizes using the \ac{SR} optimizer. Energy is shown in (\subref{fig:act_energy}) and variances are shown in (\subref{fig:act_variance}). }
  \label{fig:act_compare}
\end{figure*}

In the previous sections, we only probed two different active space sizes in our \ac{OO} investigations or compared the impact of various Jastrow forms. 
Here we investigate whether \ac{OO} can outperform the standard DFT+U scan of the nodal surface by systematically increasing the number of virtual orbitals included in the optimization. 
While we previously saw that performing \ac{OO} was not able to produce a lower \ac{DMC} energy for the active space sizes considered, it could be the case that the wave function is not sufficiently converged with respect to active space and larger active spaces will eventually also improve the \ac{DMC} energy. 

In Figure (\ref{fig:act_compare}), we compare the energies and variances for different active space size optimizations. 
These calculations range from around 1k to 30k optimization parameters for the calculations shown. 
Clearly, by increasing the active space size, there is substantial energy gain.
Even once 300 orbitals is reached, it seems that lower VMC energies could be obtained with larger active spaces. 
However, the variance is less sensitive to active space size once there are enough virtual orbitals. 

\begin{figure*}
  \centering
  \begin{subfigure}[b]{0.48\textwidth}
    \centering
    \includegraphics[width=\linewidth]{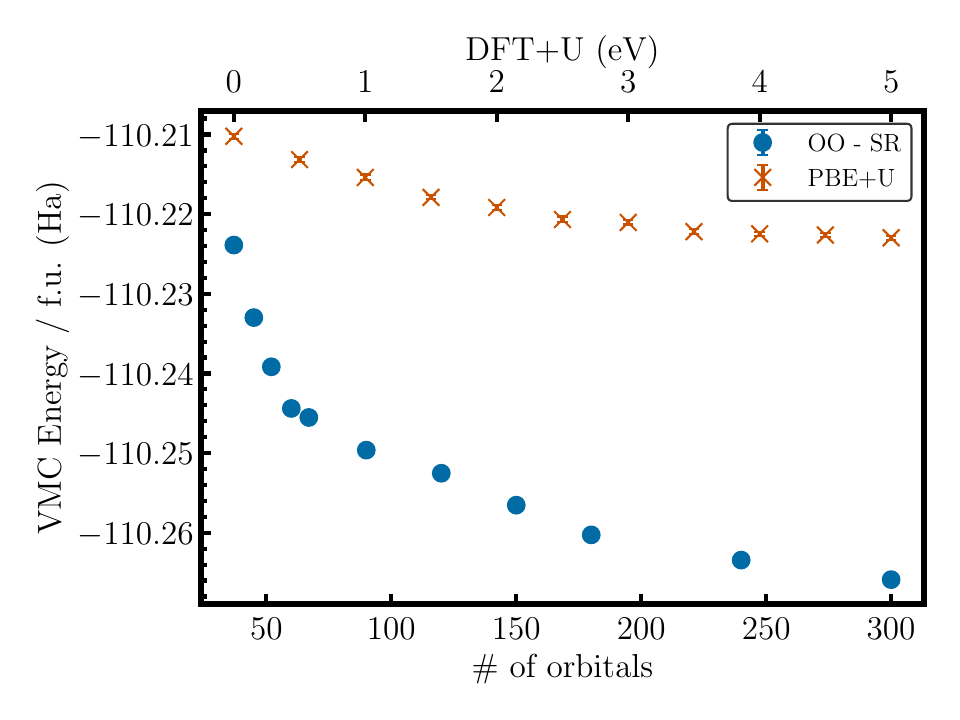}
    \caption{\ac{VMC} energies}
    \label{fig:act_vmc}
  \end{subfigure}
  \hfill
  \begin{subfigure}[b]{0.48\textwidth}
    \centering
    \includegraphics[width=\linewidth]{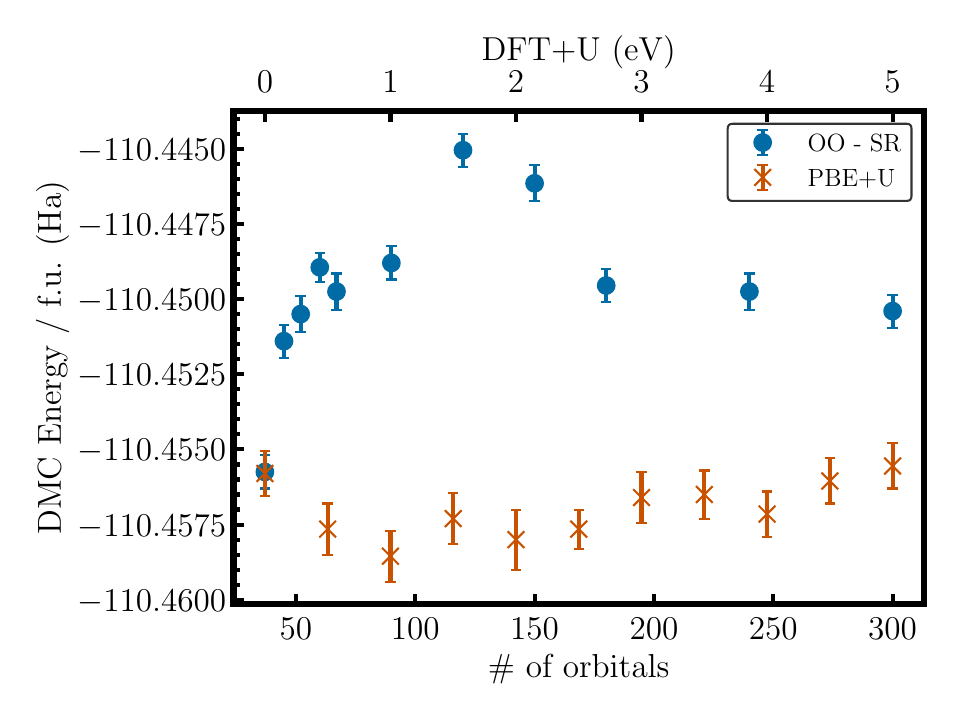}
    \caption{\ac{DMC} energies}
    \label{fig:act_dmc}
  \end{subfigure}
  \caption{Comparison of \ac{OO} using \ac{SR} with various active space sizes, compared to a baseline of DFT+U orbitals with only a scan of the U parameter. }
  \label{fig:act_qmc}
\end{figure*}

Given the current active space sizes and convergence, we carried out \ac{VMC} and \ac{DMC} calculations and now compare again to the simplest optimization, i.e. scanning the DFT+U orbitals. 
The results are shown in Figure \ref{fig:act_qmc}.
At the \ac{VMC} level, we significantly outperform the \ac{DFT}+U orbitals and clearly have a better \ac{SDJ} variational ansatz.  In VMC, the energy decreases monotonically with increasing active space size.
The \ac{DMC} energies, however, behave differently. There the initial \ac{DMC} energies get worse with active space size, and then eventually start to improve again around 120 orbitals (or 10k parameters). 
The energies saturate in quality by around 180 orbitals and remain consistently higher than the \ac{DFT}+U by $\sim$6~mHa.
It should be noted that this difference is similar in magnitude to the DMC energy gained when using J123 instead of J12 in the optimization, as shown in Figure \ref{fig:jastrow_scan}.  
Therefore it is not unreasonable to expect that improved short-ranged Jastrow forms could result in better energetics that do not underperform relative to DFT orbitals.

\begin{figure}
    \centering
    \includegraphics[width=\linewidth]{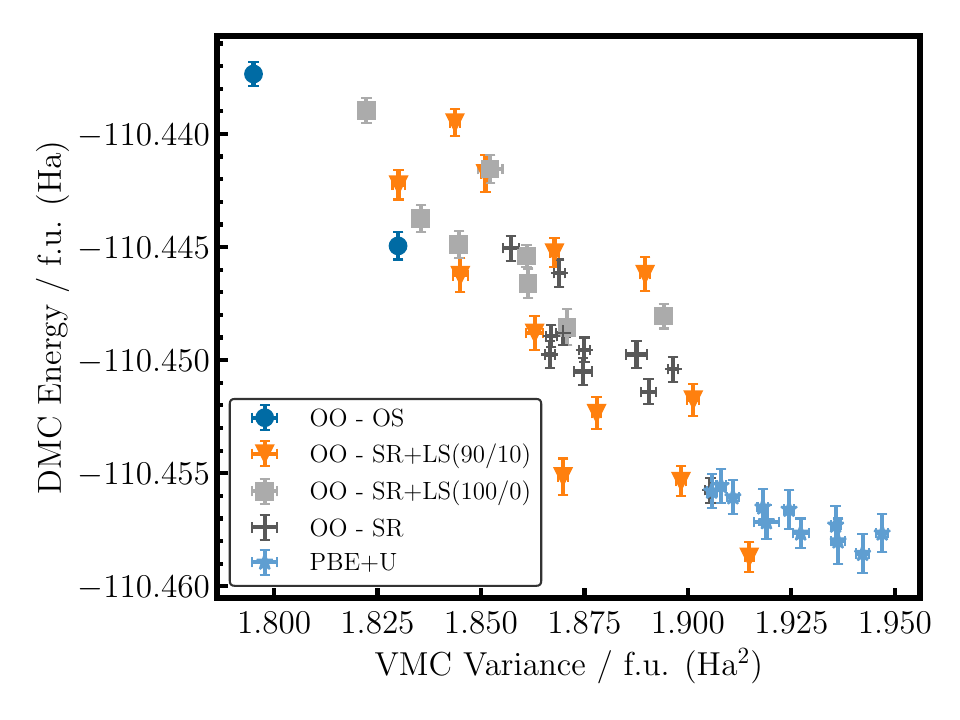}
    \caption{\ac{DMC} energy vs. \ac{VMC} variance across a wide range of active spaces and different optimizers. }
    \label{fig:dmcen_vmcvar}
\end{figure}

It is also interesting to examine how the DMC energy and VMC variance are related, as shown in Figure \ref{fig:dmcen_vmcvar}.
Somewhat counter intuitively, for the \ac{SDJ} form, wave functions with the lowest \ac{DMC} energy are those with the highest \ac{VMC} variance.  This trends holds clearly across all OO wavefunctions, regardless of the active space size or the employed optimization method. 
The DFT+U orbitals result in the highest VMC variances and lowest \ac{DMC} energies of all.  Interestingly, using the pure SR method consistently gives higher variance than all other optimization approaches, while also giving the lowest \ac{DMC} energies among the \ac{OO} wave functions.

As we have already described, we suffer from both fixed node error as well as localization error of the pseudopotential. 
It is possible that the \ac{OO} procedure has actually improved the fixed-node error but has increased the locality error sufficiently to raise the \ac{DMC} energy overall, vice-versa, or increased both errors. 
As the Jastrow functions are systematically improved to approach the true fixed-node wave function, the locality error vanishes and the \ac{VMC} energy would agree with the \ac{DMC} energy \cite{krogel_magnitude_2017}. 

\begin{figure}
    \centering
    \includegraphics[width=\linewidth]{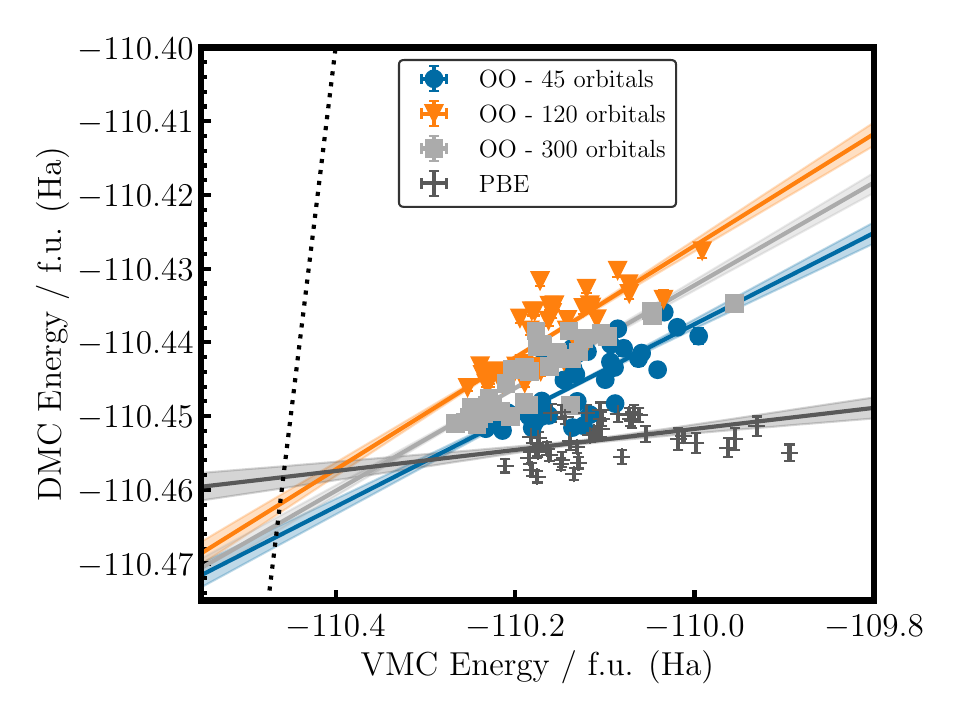}
    \caption{Estimation of pure fixed-node errors for multiple orbital sets, including pure PBE as well as \ac{OO} with various active spaces. The intersection of the various fits with the $E_{\rm DMC} = E_{\rm VMC}$ line indicates the pure fixed-node energy.  The points represent wave functions with differing Jastrow factors.   }
    \label{fig:locality}
\end{figure}

Therefore, we can exploit this fact to estimate the pure fixed-node error through extrapolation using multiple Jastrow factors, shown in Fig. \ref{fig:locality}
For each orbital set, each point corresponds to a new Jastrow factor, where we randomly perturb each parameter up to a 5\% change to generate a new Jastrow. 
Then both \ac{VMC} and \ac{DMC} calculations are performed with each of those new Jastrows with fixed orbitals. 
We consider the unoptimized PBE orbitals, as well as various \ac{OO} orbital sets with varying active space size and perform linear fits through the data. 
The line where $E_{\rm DMC} = E_{\rm VMC}$ represents the theoretical condition where there is no locality error and the remaining bias is due to fixed node error. 
From this, it is clear that \ac{OO} is actually improving the fixed-node bias since we intersect the $E_{\rm VMC} = E_{\rm DMC}$ at lower energies. 
This is true for all active spaces, although 120 orbitals shows the largest fixed-node bias among the \ac{OO} calculations. 

The magnitude of the locality error for a given wave function error is given by the \ac{DMC} energy difference from the pure fixed-node energy. 
From this, we see that the locality error is systematically increased when using \ac{OO} when compared to fixed PBE orbitals. 
Taken together, \ac{OO} actually improves the fixed-node error by a few mHa over bare DFT orbitals. 
In this case, and for a currently unknown reason, it increases the locality error significantly, which reduces the accuracy of the \ac{DMC} energies, at least within the T-moves approximation. 

The overall errors relative to standard fixed orbital sets is nearly the same magnitude as the gain in \ac{DMC} energy moving from a two- to three-body Jastrow factor.
Therefore, further improvements to short-ranged Jastrow factors may be necessary in order to fully surpass the fixed PBE orbitals. 
We note that the magnitude of the locality error is largely system and pseudopotential dependent, so it is possible that \ac{OO} will show robust improvements in some systems. 
Additionally, calculations with reduced or zero locality error, such as all-electron or pseudo-Hamiltonian approaches \cite{krogel_hybridizing_2020,bennett_high_2022, ichibha_locality_2023}, may show more clear improvements from \ac{OO} alone, although this is left for future investigations by ourselves or others. 

\subsection{Mixed Estimator Bias}
\label{sec:oo_mixed_est}
In many cases we are interested in more than just the \ac{DMC} energies and also care about various observables, such as the spin density, magnetization density, one-body reduced density matrix, etc. 
In each case, these observables do not commute with the Hamiltonian and therefore have a mixed-estimator bias due to  
sampling the mixed-distribution $\Psi_T\Psi_0$. 
A simple estimate of the correct \ac{DMC} estimator is given by the extrapolated estimator approximation, i.e.
\begin{equation}
    \langle \hat{O} \rangle_{\Psi_0} \approx \langle \hat{O} \rangle_{est} = 2 \langle \hat{O}\rangle_{\rm DMC} - \langle \hat{O} \rangle_{\rm VMC}
\end{equation}
Therefore, the mixed-estimator bias can be estimated by 
\begin{equation}
    \Delta \hat{O}_{\rm mix} = \langle \hat{O}\rangle_{\rm DMC} - \langle\hat{O}\rangle_{\rm VMC}
\end{equation}
This implies that as the trial wavefunction becomes more accurate, e.g. via a lower energy or variance,  then the mixed estimator approaches the true \ac{DMC} estimator, and therefore the mixed-estimator bias will be reduced.

\begin{figure}
    \centering
    \includegraphics[width=\linewidth]{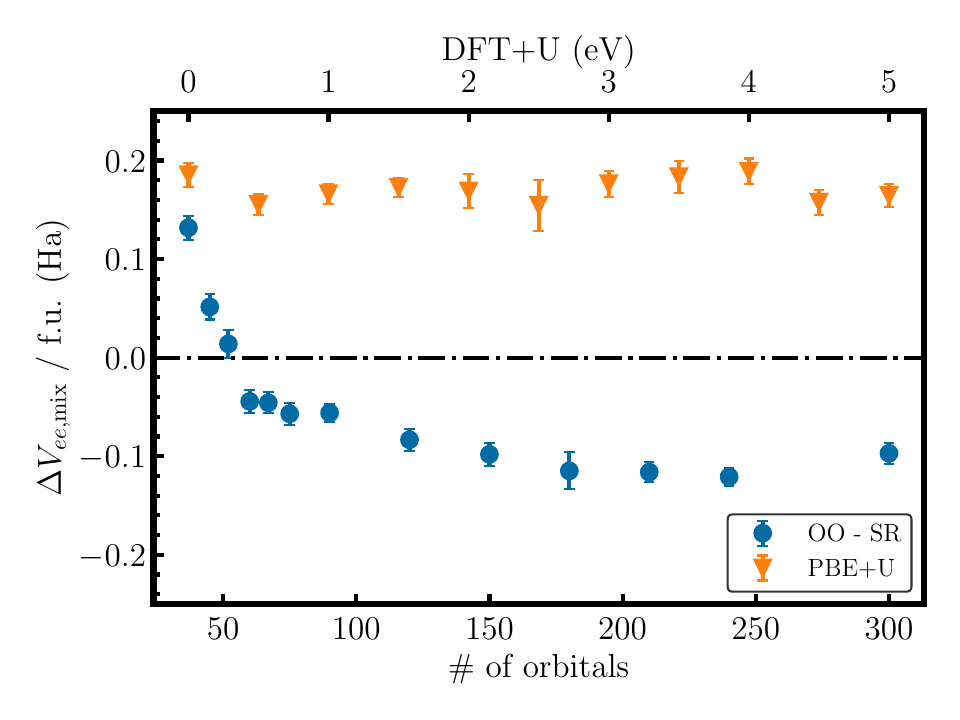}
    \caption{Mixed estimator bias of the electron-electron interaction energy compared between various active space sizes of \ac{OO} using the \ac{SR} optimizer and a standard DFT+U scan. }
    \label{fig:mixed_ee}
\end{figure}

As seen above, we are in a regime where DFT+U trial wave functions give the lowest \ac{DMC} energies, but significantly higher \ac{VMC} energies. 
Therefore, variationally the \ac{OO} wavefunctions are better and potentially result in a lower mixed-estimator bias even if there is a small compromise in the \ac{DMC} energy.
As a check, we show the mixed-estimator bias of the electron-electron interaction energy in Figure \ref{fig:mixed_ee}. 
The electron-electron interaction energy is a useful proxy for other non-commuting observables, as it is closely tied to the electron density. 
By performing a DFT+U scan, the mixed-estimator bias is roughly constant across all values of U and is around 0.2~Ha/f.u. 
By performing \ac{OO}, we see a significant reduction in the mixed estimator bias that improves with active space size. 
Even though the \ac{DMC} energies are slightly worse in this case using \ac{OO}, we see a very clear improvement in the mixed estimator bias that is due largely to the significantly improved variational wave function used in \ac{VMC}.

\begin{figure*}
  \centering
  \begin{subfigure}[b]{0.48\textwidth}
    \centering
    \includegraphics[width=\linewidth]{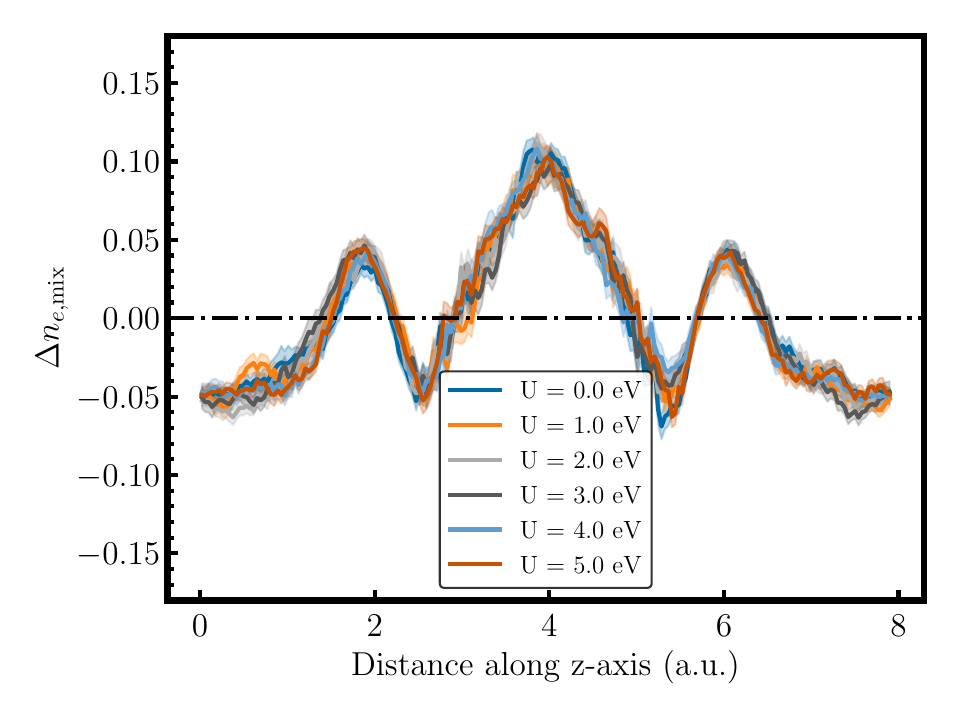}
    \caption{PBE+U wave functions}
    \label{fig:mixed_den_pbeu}
  \end{subfigure}
  \hfill
  \begin{subfigure}[b]{0.48\textwidth}
    \centering
    \includegraphics[width=\linewidth]{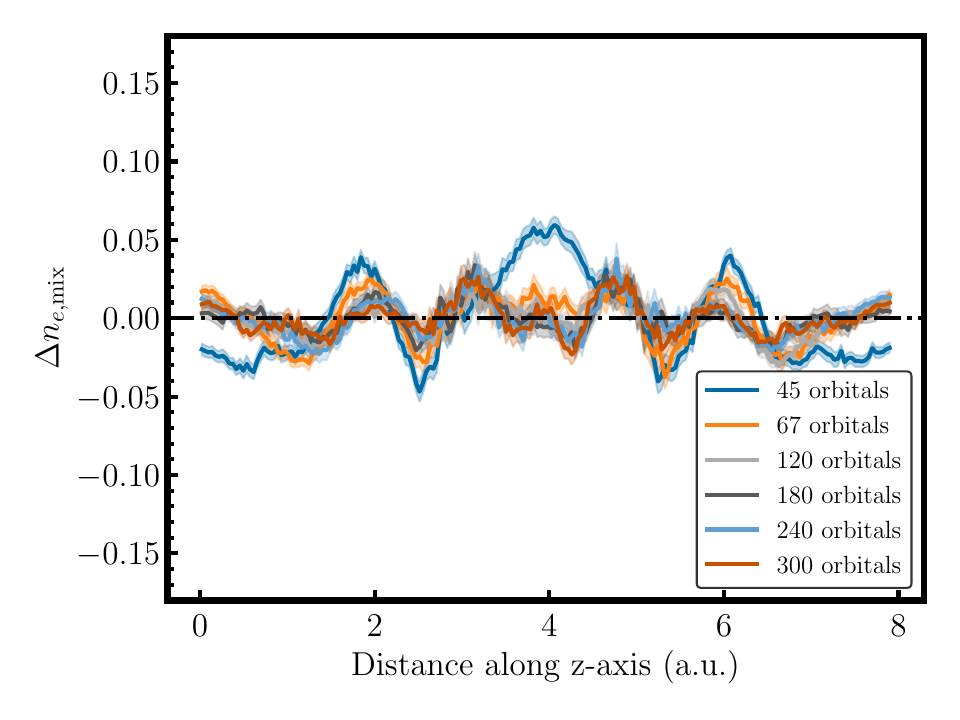}
    \caption{\ac{OO} wave functions}
    \label{fig:mixed_den_oo}
  \end{subfigure}
  \caption{Comparison of \ac{OO} using \ac{SR} with various active space sizes, compared to a baseline of DFT+U orbitals with only a scan of the U parameter for the mixed estimator bias of the electron density. }
  \label{fig:mixed_den}
\end{figure*}

We also analyze the mixed-estimator bias for the electron density estimator. 
We show the mixed estimator bias of the density along the $z$ axis of the cell (with $x,y$ = 0) for both \ac{DFT}+U wave functions and \ac{OO} wave functions in Fig. \ref{fig:mixed_den}. 
In this case, the bias using PBE+U orbitals is relatively insensitive to the choice of +U. 
\ac{OO} for all active space size results in a smaller bias for the density, which decreases as the quality of the \ac{OO} increases. 

\section{Conclusions}
\label{sec:conclusions}

In this work we investigated the impact of \ac{OO} in the correlated magnetic insulator CrSBr from a variety of angles. 
First we considered the performance of various optimizers available within \textsc{QMCPACK}.
Although each method has its roots in minimizing the variational energy, we find the optimizers produce qualitatively different wave functions, with pure \ac{SR} giving the lowest variational energy despite a slower convergence rate. 
We speculate that the use of correlated sampling biases the other optimizers, but further investigation is required. 
We also probe the coupling between Jastrow quality and nodal surfaces by optimizing various long and short range Jastrows simultaneously with the orbital rotations. 
At the \ac{VMC} level, \ac{OO} robustly produces lower energies than fixed \ac{DFT}(+U) orbitals for all Jastrow factors. 
Within \ac{DMC}, improvements to the short range Jastrows systematically improves the energy for all orbital treatments, while the impact of long-range Jastrows is negligible. 
For CrSBr, large active spaces are required to converge the \ac{OO}. 
For all active spaces considered, \ac{OO} monotonically lowers the energy and is systematically lower than the standard \ac{DFT}+U orbitals scan. 
Despite the clear variational improvement, \ac{DMC} on \ac{OO} wave functions is systematically higher than the standard \ac{DFT}+U scan for all active spaces, with intermediate active spaces being the worst. 
Given the large improvement in \ac{VMC} of almost $\sim$50 mHa, the bias incurred at the \ac{DMC} level is only $\sim$ 6 mHa. 
Interestingly, we observe a strong correlation between the \ac{VMC} variance and \ac{DMC} energy, with larger variance showing the best \ac{DMC} energies for any orbital treatment. 
By performing energy extrapolations using Jastrow factor variations, we show that \ac{OO} is generating better nodes than \ac{DFT}+U. 
However, there is a systematic increase in the locality error associated with \ac{OO}, resulting in an overall higher \ac{DMC} energy. 
We also show there is a systematic improvement in mixed-estimator biases by using \ac{OO}, which is important for observables other than the energy such as the spin-density or one-body reduced density matrix \cite{shin_dft_2024,lopez_identifying_2025}. 

Further investigations are needed to discern whether \ac{VMC} improvements of the orbitals increases the \ac{DMC} locality error is general, and to uncover the cause. 
While there is no guarantee that directly improving the \ac{VMC} energy through optimization should also improve the \ac{DMC} energy, this has been shown for more flexible neural network wave functions, without pseudopotentials and locality errors, which are only a few mHa away from the exact energies, i.e. Fig. 3b in reference \cite{ren_towards_2023}. 
Seemingly if the variational wave function is sufficiently flexible, improved \ac{VMC} energies do yield to improved \ac{DMC} energies. 
Our results suggest the fixed node energy is indeed improving even for simple Slater Jastrow despite higher locality error. 
Because of this, we would expect to see improved \ac{DMC} energies for calculations that avoid locality errors, e.g. all-electron or pseudo-Hamiltonian approaches \cite{krogel_hybridizing_2020,bennett_high_2022, ichibha_locality_2023}. 
Regardless, \ac{OO} is a promising and scalable approach to improving the variational energy, fixed-node error, and mixed-estimator biases for arbitrary observables at the cost of a small increase in locality error. 

\section{Acknowledgments}
C.A.M. would like to thank Raymond C. Clay III, Amanda Dumi, and Joshua P. Townsend for many useful discussions. 

This work was supported by the U.S. Department of Energy, Office of Science, Basic Energy Sciences, Materials Sciences and Engineering Division, as part of the Computational Materials Sciences Program and Center for Predictive Simulation of Functional Materials.

Sandia National Laboratories is a multi-mission laboratory managed and operated by National Technology \& Engineering Solutions of Sandia, LLC (NTESS), a wholly owned subsidiary of Honeywell International Inc., for the U.S. Department of Energy’s National Nuclear Security Administration (DOE/NNSA) under contract DE-NA0003525. This written work is authored by an employee of NTESS. The employee, not NTESS, owns the right, title and interest in and to the written work and is responsible for its contents. Any subjective views or opinions that might be expressed in the written work do not necessarily represent the views of the U.S. Government. The publisher acknowledges that the U.S. Government retains a non-exclusive, paid-up, irrevocable, world-wide license to publish or reproduce the published form of this written work or allow others to do so, for U.S. Government purposes. The DOE will provide public access to results of federally sponsored research in accordance with the DOE Public Access Plan. 

This manuscript has been authored by UT-Battelle, LLC under Contract No. DE-AC05-00OR22725 with the U.S. Department of Energy. The United States Government retains and the publisher, by accepting the article for publication, acknowledges that the United States Government retains a non-exclusive, paid-up, irrevocable, worldwide license to publish or reproduce the published form of this manuscript, or allow others to do so, for United States Government purposes. The Department of Energy will provide public access to these results of federally sponsored research in accordance with the DOE Public Access Plan (http://energy.gov/downloads/doe-public-access-plan).


\section*{Data Availability}
The data that supports the findings of this study are available within the article and its supplementary material.

\section*{Author Declarations}
The authors have no conflicts to disclose.

\bibliography{references}

\end{document}